# Measuring Team Creativity Through Longitudinal Social Signals

Peter A. Gloor, Adam Almozlino, Orr Inbar
*MIT Center for Collective Intelligence*
*5 Cambridge Center, Cambridge MA 02139, USA*
*pgloor@mit.edu*

Wei Lo
*Computer Science and Technology department*
*Zhejiang University, Hangzhou, P.R. China*

Shannon Provost
*McCombs School of Business, University of Texas at Austin*
*Austin, TX, USA*

**Summary**

*Research into human dynamical systems has long sought to identify robust signals for human behavior. We have discovered a series of social network-based indicators that are reliable predictors of team creativity and collaborative innovation. We extract these signals from electronic records of interpersonal interactions, including e-mail, and face-to-face interaction measured via sociometric badges. The first of these signals is Rotating Leadership, measuring the degree to which, over time, actors in a team vary in how central they are to team's communication network's structure. The second is Rotating Contribution, which measures the degree to which, over time, actors in a team vary in the ratio of communications they distribute versus receive. The third is Prompt Response Time, which measures, over time, the responsiveness of actors to one another's communications. Finally, we demonstrate the predictive utility of these signals in a variety of contexts, showing them to be robust to various methods of evaluating innovation.*

**Introduction**

In this paper we introduce a series of longitudinal, network-based measures of social interaction patterns that predict collaborative innovation. Innovation is a universal, emergent human behavior. According to noted evolutionary biologist E.O Wilson "…it was necessary for the evolving populations to acquire an ever higher degree of intelligence. They had to feel empathy for others, to measure the emotions of friends and enemy alike, to judge the intentions of all of them, and to plan a strategy for personal social interactions" [1]. Innovation is a universal, emergent human behavior, one that rarely occurs through the actions of a single individual, but rather through collaboration among individuals [2]. Here we focus on the predictive utility of observing this collaboration at the level of interpersonal interaction events.

Recently, researchers have had success in identifying reliable, quantitative indicators of phenomena in human systems. Among these indicators are "honest signals" [3][4][5], which signify the presence of social influence. This name captures both the separation of these signals' from the subjectivity that often plagues other methods for measuring human behavior, and the robustness of these signals to a variety of behavioral contexts. Understanding these "honest signals" can convey a significant advantage. To quote E.O Wilson again, "…social intelligence was therefore always at a high premium. A sharp sense of empathy can make a huge difference and with it an ability to manipulate, to gain cooperation, and to deceive" [1]. Robust, quantitative measures for collective human behavior may serve as the quantitative, larger-scale analog for individual social intelligence.

Previous work studying collective creativity and innovation has been primarily qualitative, focused on the creativity of individuals, or both [6][7]. Other research has been restricted to a particular interpretation of creativity, studying for example patent production [8], or to a particular setting, studying for example large corporations [9]. Therefore, this research has failed to identify reliable signals of collective innovation.

Part of the reason previous work has had limited success may lie in the difficulty of understanding innovation itself. A formal definition of innovation remains elusive, as does the boundary between incremental improvements and innovative change. If a certain dependent variable, such as creativity, is difficult to formally define, it may be difficult to identify a quantitative and reproducible independent variable that indicates the dependent.

**Our Approach**

We have attempted to work around this issue by evaluating several different proxies for creativity across several different scenarios, and identifying measures that reliably signal the presence of these proxies across the scenarios. Using a wide selection of proxies in a variety of context, we have identified reproducible independent variables that strongly correlate with the proxies. We term these variables (1) Rotating Leadership, (2) Rotating Contribution, (3) Prompt Response Time.

From these variables, Rotating Leadership and Rotating Contribution show positive correlation in "creative" work scenarios, but strong negative correlation with "non-creative" scenarios, suggesting that Rotating Leadership and Rotating Contribution are a good "honest signal" for team *creativity*. This corresponds with the intuition that creative work requires innovation and breaking known patterns of thought and behavior, while breaking known patterns may disrupt non-creative work. Prompt Response Time, on the other hand, shows positive correlation across all scenarios, suggesting that it is a better indicator of team *productivity*. This corresponds with the intuition that it is, in general, better to have a more promptly communicating team.

**(1) Rotating Leadership (RL)**

Rotating Leadership (RL) measures the degree to which, over time, the members in a team vary in how "central" they are to the team's communications**.** The advantage of centralized leadership for creative tasks was for instance observed among Wikipedians [10], where it was found that Wikipedia articles authored by more centrally communicating editors became articles of the highest quality (featured articles) more rapidly. RL can be observed in a visualization of a network when distinct nodes appear, over time, to oscillate between central and peripheral positions in the network. Intuitively, RL evaluates how much, across time and the team members, team members switch between being highly central to the overall communications of the team, and being peripheral to those communications. Formally, RL measures oscillations in Betweenness Centrality (BC) over time among actors in the team.

The effects of the centrality of team's actors to the team's performance was first observed among teams of Eclipse open source developers communicating electronically [12]. It was subsequently observed in a study of a marketing team in a bank communicating face-to-face [13], and in a study of nurses communicating in a hospital [14]. In this last scenario, quantitative measures were compared with personality characteristics such as openness, as measured by the Neo-FFI [15], and group creativity was measured through peer and management/instructor assessment, based on the premise that experts can identify creativity [7]. Note that teams composed of highly intelligent individuals are not necessarily intelligent as a whole [16], while measures such as RL were dependably correlated with team creativity.

Betweenness centrality [11] (BC) is a global measure of how centrally located a node is in the structure of a network. For a given node, it is measured by evaluating the shortest paths in the network, specifically, the proportion of all shortest paths in the network that pass through the node of interest. Mathematically, BC of a node $v$ is defined as:

$$g(v) = \sum_{s \neq v \neq t} \frac{\sigma_{st}(v)}{\sigma_{st}}$$

where $\sigma_{st}$ is the total number of shortest paths from node $s$ to node $t$ and $\sigma_{st}(v)$ is the number of those paths that pass through $v$.

In order to calculate RL, it is necessary to aggregate measures of BC, which occur at the scale of an individual actor at an individual time step, to the scale of the whole network over the full time frame. In order to do this in a fashion that indicates variation in BC we counted the number of local maxima and minima in the vector of each actor's BC over time, and then summed this number across the actors in a team.

Formally, we count the local maxima of function $f(t)=g(t)$ within time interval $[t_1,t_2]$. There is a local maximum for time t at point $t^*$, if there exists some $\varepsilon > 0$ such that $f(t^*) \geq f(t)$ when $|t - t^*| < \varepsilon$. Similarly, we count the local minima at $t^*$, if $f(t^*) \leq f(t)$ when $|t - t^*| < \varepsilon$.
RL for actor i over time window ws is therefore:

$$RL_i = \#local\ minima_i^{ws} + \#local\ maxima_i^{ws}$$

$$RL = \frac{1}{n} \sum_{i=1}^{n} RL_i$$

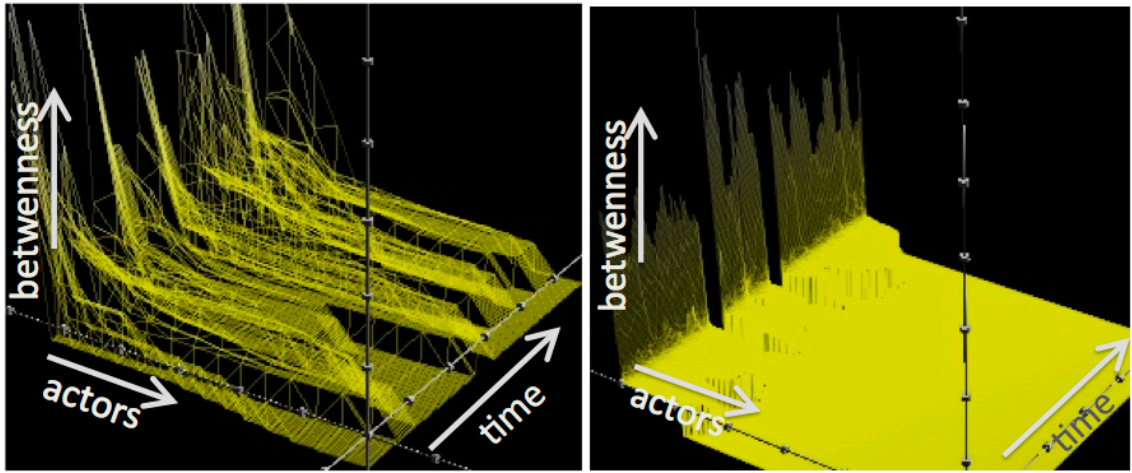

**Figure 1: RL visualized through oscillations in BC over time [17]**

*This figure illustrates Rotating Leadership (RL) for two teams, one with high RL, and one with low RL. Actors are placed along the Y-axis, while the X-axis encodes time, and the Z-axis the Betweenness Centrality (BC) of actors for each hour, sorted, each hour, by the decreasing BC of actors. The back plane, which rises and falls, represents the set of actors who rotate taking the lead in the team's communication.*

*The left picture illustrates an example of a team with high RL. This example was drawn from a 6-day long graduate student seminar, and communications were measured using sociometric badges. This image includes 15 actors, and has had BC oscillation computed every hour using a time window of 12 hours, with a date range 6/13/2010 12:37 pm to 6/19/2010 23:37 pm.*

*The right picture illustrates an example of a team with low RL. This example was drawn from the customers and employees of a large service provider serving one customer, and communications were measured using the email archive of the service provider. This image includes 2857 actors, and has had BC oscillation computed every day using a time window of 7 days, with a date range between 6/13/2012 to 12/30/2012. The high back represents the key account managers who are consistently taking the lead in team communication.*

### (2) Rotating Contribution (RC)

Rotating Contribution (RC) measures the degree to which, over time, actors in a team vary in how much they broadcast communications versus listen to communications. RC can be observed in a visualization of a network when distinct nodes appear, over time, to vary widely in how many incoming versus outgoing links they have. Intuitively, RC evaluates how much, across time and the team members, team members switch off between broadcasting many communications and listening to may communications. Formally, RC measures oscillations, over time, of the Contribution Index (CI) of actors in a team.

Contribution Index (CI) is a measure of how much an actor disseminates versus receives communications. For a given node, it is equal to ratio of incoming to outgoing links incident upon that node. An actor that only sends messages will have a CI of 1, an actor that sends and receives an identical number of messages will have a CI of 0, and an actor that only receives messages will have a CI of -1 [18]. Formally, the CI of an actor over a given time frame is:

$$CI = \frac{messages\_sent - messages\_received}{messages\_sent + messages\_received}$$

In order to calculate RC, it is necessary to aggregate measures of CI, which occur at the scale of an individual actor at an individual time step, to the scale of the whole network over the full time frame. In order to do this in a fashion that indicates variation in CI we counted the number of local maxima and minima in the vector of each actor's CI over time, and then summed this number across the actors in a team.

Formally, we count the number of local maximum points of function $f(t)=c(t)$ within time interval $[t_1,t_2]$. There is a local maximum for time t at point $t^*$ if there exists some $\varepsilon > 0$ such that $f(t^*) \geq f(t)$ when $|t – t^*| < \varepsilon$. Similarly, we count the local minima at $t^*$, if $f(t^*) \leq f(t)$ when $|t – t^*| < \varepsilon$. $RC_i^{ws}$ for actor i and time window ws is therefore

$$RC_i^{ws} = \#local\ minima_i^{ws} + \#local\ maxima_i^{ws}$$

$$RC = \frac{1}{n}\sum_{i=1}^{n} RC_i$$

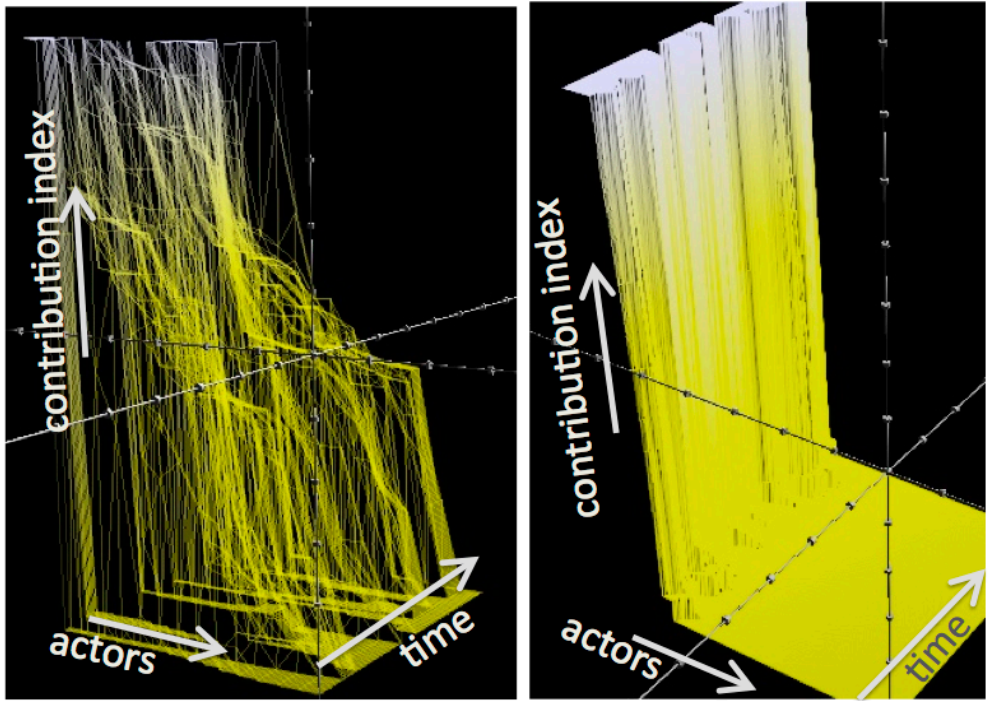

Figure 2: RC visualizing though CI oscillations over time [17]

*This figure illustrates Rotating Contribution (RC) for two teams, one with high RC, and one with low RC. Actors are placed along the Y-axis, while the X-axis encodes time, and the Z-axis the Contribution Index (CI) of actors for each hour, sorted, each hour, by the decreasing CI of actors. The back plane, which rises and falls, represents the set of actors who rotate taking the lead as most vocal contributors.*
*The left picture illustrates an example of a team with high RC; RC oscillates highly among time steps and the actors of the team. This example was drawn from a 6-day long graduate student seminar, and communications were measured using sociometric badges. This image includes 15 actors, and has had*

*BC oscillation computed every hour using a time window of 12 hours, with a date range 6/13/2010 12:37 pm to 6/19/2010 23:37 pm.*
*The right picture illustrates an example of a team with low RC; CI oscillates relatively little among time steps and the actors of the team. This example was drawn from the customers and employees of a large service provider serving one customer, and communications were measured using the email archive of the service provider. This image includes 2857 actors, and has had CI oscillation computed every day using a time window of 7 days, with a date range between 6/13/2012 to 12/30/2012. The high back represents the key account managers who are consistently the most vocal by sending more emails than they receive.*

### (3) Prompt Response Time (PRT)

Prompt Response Time (PRT) measures the degree to which, over time, actors are prompt at communicating to those who have communicated to them. Intuitively, PRT corresponds with how fast, across actors in a network, actors are at "getting back" to each other's communications. In order to capture this formally, PRT is defined in terms of the Communication Frame (CF), which groups communication events between pairs of actors which may "get back" to each other, and Frame Nudges, which measure the number of communication events in a CF, and Elapsed Time, which measures the time duration of a CF.

A *Communication Frame* (CF) groups a set of time-adjacent communications between a pair of actors. Suppose a pair of actors X and Y in a network, with a set of communication events, or directional, time-stamped edges, between them. A single CF defines all communication events from X to Y, prior to and including a communication event from Y to X. Intuitively, this is all the messages your colleague has sent you since you last messaged her, prior to and including the first message you send back to your colleague. In this framework, you, actor X, are the "source" actor in the CF, corresponding with the origin of the first communication in the CF, and your colleague, actor Y, is the "target" actor in the CF, corresponding with the origin of the last communication in the CF. The Elapsed Time (ET) for this CF is the difference between the first communication in the frame and the last communication in the frame. The Frame Nudges (FN) for this CF is the number of communication events in the CF, intuitively FN is the number of "pings" X sends until Y responds.

To get the network-level measure of PRT from the edge-level measure of ET in CFs it is necessary to aggregate measures. We accomplished this by using an intermediate actor-scale measure, where we evaluated the "responsiveness" of actors through their Responsiveness in Communication Frames (RCF). Intuitively, we measure how quickly actors got back to people who messaged them.

This can be accomplished either by measuring the ET or the FN of CFs. We define RCF via ET (RCF-ET) for an actor as the mean ET for all CFs in which the node is the "target" node. We define RFC via FN (RCF-FN) for an actor as the mean FN for all CFs in which the node is the "target" node. For actor i, where $f$ is a given CF in the set of CF denoted as $f \in F$, and $\Delta T_f$ is the time elapsed for frame $f^*$, such that $f^* \in F \cap i$, where $F \cap i$ represents the set of all frames that *actor i* is a member, RCF -ET is:

$$RCF\text{-}ET_i = \frac{\sum_{f=1}^{|F \cap i|} \Delta T_f}{F \cap i}$$

For actor i, where $f$ is a given CF in the set of CF denoted as $f \in F$, and FN is the number of edges in frame $f^*$, such that $f^* \in F \cap i$, where $F \cap i$ represents the set of all frames that *actor i* is a member, RFC-FN is:

$$RCF\text{-}FN_i = \frac{\sum_{f=1}^{|F \cap i|} FN_f}{F \cap i}$$

We then aggregate this actor-scale measure to the networks-scale by averaging RCF for all actors in the network. This procedure is the same for RCF-ET and RCF-FN. For a network, where $n_i$ denotes the number of communications of each actor $i$, PRT is therefore:

$$\text{PRT} = \frac{\sum_{i=1}^{N} RCF_i * n_i}{\sum_{i=1}^{N} n_i}$$

**Analysis and Results**

We extracted signals of team creativity and productivity from electronic records of interpersonal interactions, including e-mail, and face-to-face interaction measured via sociometric badges [28]. Some of our samples have quite a small N (<10) because of the difficulty of obtaining the type of small group communication data we are analyzing - small team communication networks which are associated with a measure of creativity and/or performance. This is compensated for by the comparability of the 5 datasets that allow for cross-comparative validation across a wide range of small group and larger organizational settings. For each scenario, we measured the Rotating Leadership, Rotating Contribution, and Prompt Response Time measured by Elapsed Time and Frame Nudges, for the teams recorded (Table 1). We studied the following five scenarios, captured via the described datasets:

|  | Dependent variable | Interaction type | #actors | #interaction records | duration |
|---|---|---|---|---|---|
| Global Virtual Course | creativity | e-mail | 161 | 3782 | 3 months |
| Co-located Course | creativity | sociometric badges | 15 | 265,160 | 5 days |
| Eclipse developers | Creativity, performance | mailing list | 1371 | 6405 | 6 months |
| Medical researchers | creativity | e-mail | 22,523 | 117,027 | 12 months |
| Service Provider | performance | e-mail | 85,680 | 7,640,016 | 7 months |

*Table 1. Basic parameters of 5 datasets employed to verify "honest signals"*

(1) (COINscourse2012 [19]) – An e-mail archive of a multinational, distributed graduate student seminar. Contains 161 actors and 3782 messages. 50 students were divided into 10 student teams at five in three countries on two continents. These students worked together as distributed virtual teams over 14 weeks. The dependent variable for creativity was taken as the mean of peer-ratings of students, and from an instructor rating.

(2) (CGSseminar2010 [20]) – A face-to-face interaction archive of a co-located course, gathered through sociometric badges at a doctoral seminar. Contains 15 participants, who worked on different projects in nine teams during one week. The dependent variable for creativity was measured through peer ratings from participants.

(3) (Eclipse2005 [12]) - A mailing list archive of 26 working groups of Eclipse open source developers. Contains 1371 actors and 6405 messages over a period of six months. The dependent variables for performance and creativity were measured as the volume of bugs fixed (normalized by team size) and the volume of new features (normalized by of the count of fixed bugs), respectively.

(4) (ChronicCareTeams2012 [21]) – An e-mail archive with a core team of 30 clinicians and health services researchers. Contains 22,523 different actors and 117,027 messages, working on 10 different medical innovations over the period of one year. The dependent variable for creativity was measured through ratings from a senior project management team.

(5) (ServiceProvider2012 [22]) – An e-mail archive of staff members working in 14 different accounts at a global service provider. Contains 85,680 actors and 7,640,016 different messages, with an account manager coordinating activities per corporate customer. The dependent variable for customer satisfaction was measured through a Net Promoter Score [23].

| COINscourse2012 | | RL | RC | PRT - FN | PRT - ET |
|---|---|---|---|---|---|
| creativity | Pearson Correlation | **.830**\*\* | **.928**\*\* | **.796**\*\* | -.610 |
| | Sig. (2-tailed) | .003 | .000 | .006 | .061 |
| | N | 10 | 10 | 10 | 10 |
| **CGSseminar2010** | | | | | |
| creativity | Pearson Correlation | **.707**\* | **.733**\* | .368 | .275 |
| | Sig. (2-tailed) | .033 | .025 | .370 | .509 |
| | N | 9 | 9 | 8 | 8 |
| quality | Pearson Correlation | .277 | .261 | **.882**\*\* | **.954**\*\* |
| | Sig. (2-tailed) | .470 | .498 | .004 | .000 |
| | N | 9 | 9 | 8 | 8 |
| **Eclipse2005** | | | | | |
| bugs_fixed | Pearson Correlation | -.092 | -.200 | -.366 | **-.546**\*\* |
| | Sig. (2-tailed) | .654 | .328 | .078 | .006 |
| | N | 26 | 26 | 24 | 24 |
| performance | Pearson Correlation | **-.754**\*\* | **-.698**\*\* | -.266 | -.161 |
| | Sig. (2-tailed) | .000 | .000 | .220 | .462 |
| | N | 25 | 25 | 23 | 23 |
| creativity | Pearson Correlation | .216 | .246 | **.554**\*\* | -.084 |
| | Sig. (2-tailed) | .289 | .226 | .005 | .697 |
| | N | 26 | 26 | 24 | 24 |
| **ChronicCareTeams2012** | | | | | |
| creative performance | Pearson Correlation | **.753**\* | **.751**\* | -.117 | .262 |
| | Sig. (2-tailed) | .012 | .012 | .749 | .465 |
| | N | 10 | 10 | 10 | 10 |
| creativity | Pearson Correlation | .231 | .287 | **.730**\* | .571 |
| | Sig. (2-tailed) | .520 | .422 | .017 | .085 |
| | N | 10 | 10 | 10 | 10 |
| **ServiceProvider2012** | | | | | |
| performance | Pearson Correlation | **-.589**\* | **-.618**\* | .429 | **.629**\* |
| | Sig. (2-tailed) | .034 | .019 | .164 | .029 |
| | N | 13 | 14 | 12 | 12 |

*Table 2. Correlations between 3 Social-Network based indicators and Creativity for the five test datasets. \*\*. Correlation is significant at the 0.01 level (2-tailed). \*. Correlation is significant at the 0.05 level (2-tailed)*

**Discussion**

Dynamic Social Network Analysis (SNA) [24] provided us a common framework across these scenarios, allowing us to extract the same measures. SNA represents people as nodes and their connections as links which together form a network. The properties of the resulting network and its entities can be studied to glean insights about the human collection represented. SNA has been previously used to study creativity [25][26]. While adding time at the actor level is not new [27], our work complements existing methods by measuring interaction over time among teams of individuals who must necessarily communicate, allowing us to measure edge-dependent features of the network as well.

A main limitation of our study is the small N (<10) of some of our samples. This is caused by the substantial effort of obtaining the type of small group communication data we are analyzing - team communication networks which are associated with measures for creativity and performance. This limitation is compensated for by the comparability of the 5 datasets, allowing for cross-comparative validation across a wide range of small group and larger organizational settings. We also hope that the far-reaching insights into human creativity possible through the approach proposed in this paper will motivate other researchers to conduct similar studies, thus increasing the availability of data to validate and extend our approach.

Rotating leadership RL and Rotating Contribution RC are a consistent indicator of creativity, we also find that for a non-creative activity such as large account management at the global service provider RL and RC maintain predictive power, however the direction of the correlation changes: for creative tasks, more is better, for non-creative tasks, less rotation in leadership and contribution is better. The number of nudges PRT-FN is a predictor of high creativity, while – counterintuitively – taking more time (PRT-ET) for a reply leads to more satisfied customers of the service provider. PRT-ET is positively correlated to the speed of fixing software bugs, which makes intuitive sense: the faster developers answer, the faster they will also be in fixing bugs.

While the results presented are preliminary, they nevertheless illustrate that "honest signals" of communication among team members predict the creativity and performance of the team. They are therefore a first step towards defining a new science of collaboration, that delivers a novel way to measure and even optimize creativity and performance of teams by coming up with recommendations for increased communication. While the definition of "creativity" remains elusive, we have introduced a set of robust dependent metrics that have the power to predict if humans working together in a team might be engaged in a creative task.

**References**

1. Wilson EO. (2012) The Social Conquest of Earth, Liveright Publishing Corp. New York. p. 17 and p 44.
2. Sawyer C. (2007) *Group genius: the creative power of collaboration*. Basic Books, New York.
3. Pentland A. (2008) *Honest signals, How they shape our world*. MIT Press, Cambridge
4. Gloor P, Oster D, Raz O, Pentland A, Schoder D. (2010) The virtual mirror—reflecting on your social and psychological self to increase organizational creativity. *Journal International Studies of Management and Organization*. 40(2) 74-94.
5. Gloor P, Fischbach K, Fuehres H, Lassenius C, Niinimäki T, Olguin Olguin D, Pentland A, Piri A, Putzke J. (2010) Towards "Honest Signals" of Creativity – Identifying Personality Characteristics Through Microscopic Social Network Analysis. *Procedia - Social and Behavioral Sciences*, **26**. Proceedings COINs 2010, Collaborative Innovations Networks Conference, Savannah GA, Oct 7-9.
6. Csíkszentmihályi M. (1996) Creativity: flow and the psychology of discovery and invention. Harper Perennial, New York.
7. Amabile TM. (1983) *The social psychology of creativity*. Springer-Verlag, New York.
8. Jung DI, Chow C, Wu A. (2003) The role of transformational leadership in enhancing organizational innovation: Hypotheses and some preliminary findings, *The Leadership Quarterly*, 14(4–5), 525-544 August–October.
9. Woodman RW, Sawyer JE, Griffin RW. (1993) Toward a Theory of Organizational Creativity, *The Academy of Management Review*, Vol. 18, No. 2, pp. 293-321.
10. Nemoto K, Gloor P. Laubacher R. (2011) Social Capital Increases Efficiency of Collaboration among Wikipedia Editors, ACM Hypertext 2011: 22nd ACM Conference on Hypertext and Hypermedia, Eindhoven, NL, June 6-9.
11. Freeman L. (1977) A set of measures of centrality based on betweenness. Sociometry **40** 35–41.
12. Kidane Y. Gloor P. (2007) Correlating temporal communication patterns of the Eclipse open source community with performance and creativity, *Computational & Mathematical Organization Theory*. 13 (1). 17-27.
13. Fischbach K, Gloor P. Schoder D. (2009) Analysis of informal communication networks – A case study. *Business & Information Systems Engineering* **2**, February.
14. Olguin Olguin D. Gloor P. Pentland A. (2009) Capturing Individual and Group Behavior with Wearable Sensors. Proc AAAI 2009 Spring Symposium, Stanford, March 23-25.
15. McCrae RR, Costa PT, Martin TA. (2005) The NEO-PI-3: A more readable revised NEO personality inventory. *Journal of Personality Assessment*. 84 (3): 261–270.

16. Woolley AW, Chabris CF, Pentland A, Hashmi N, Malone TW. (2010) Evidence for a Collective Intelligence Factor in the Performance of Human Groups. *Science*. 29 October. **330** (6004) 686-688.
17. Gloor P. (2005) Capturing Team Dynamics Through Temporal Social Surfaces , Proceedings of 9th International Conference on Information Visualisation IV05, London, 6-8 July.
18. Gloor P. Laubacher R. Dynes S. Zhao Y. (2003) Visualization of Communication Patterns in Collaborative Innovation Networks: Analysis of some W3C working groups. ACM CIKM International Conference on Information and Knowledge Management, New Orleans, Nov 3-8.
19. Gloor P, Paasivaara M. (2013) COINs Change Leaders - Lessons Learned from a Distributed Course. Proceedings 4rd Intl. Conf on Collaborative Innovation Networks COINs 2013, Aug. 11-13, Santiago de Chile.
20. Gloor P, Grippa F, Putzke J, Lassenius C, Fuehres H, Fischbach K, Schoder D. (2013) Measuring Social Capital in Creative Teams Through Sociometric Sensors. Int. J. Organisational Design and Engineering, **2** 4.
21. Zhang X. Gloor P. Grippa F. (2013) Measuring Creative Performance of Teams Through Dynamic Semantic Social Network Analysis. Int. J. Organisational Design and Engineering, **3** 2.
22. Brunnberg D, Gloor P, Giacomelli G. (2013) Predicting Customer Satisfaction through (E-Mail) Network Analysis: The Communication Score Card. Proceedings 4rd Intl. Conf on Collaborative Innovation Networks COINs 2013, Aug. 11-13, Santiago de Chile.
23. Reichheld F. (2013) One Number You Need to Grow, Harvard Business Review. December.
24. Carley KM. (2003) Dynamic Network Analysis, in Dynamic Social Network Modeling and Analysis: Workshop Summary and Papers, Ronald Breiger, Kathleen Carley, and Philippa Pattison, (Eds.) Committee on Human Factors, National Research Council, National Research Council. 133–145, Washington, DC.
25. Cummings JN, Kiesler S. (2008) Who collaborates successfully? Prior experience reduces collaboration barriers in distributed interdisciplinary research. Proceedings of the ACM conference on Computer-Supported Cooperative Work, (November 10-12), San Diego, CA (2008).
26. Leenders RT, van Engelen JM, Kratzer J. (2003) Virtuality, communication, and new product team creativity: a social network perspective, Journal of Engineering and Technology Management, Volume 20, Issues 1–2, June, 69-92.
27. Snijders TAB. (2005) Models for Longitudinal Network Data. Chapter 11 in P. Carrington, J. Scott, & S. Wasserman (Eds.), Models and methods in social network analysis. New York: Cambridge University Press.
28. Olguín Olguín D. (2007) Sociometric badges: wearable technology for measuring human behavior. MAS Thesis, Massachusetts Institute of Technology. Cambridge, MA, May.